\documentclass[10pt]{IEEEtran}
\IEEEoverridecommandlockouts
\AtBeginDocument{%
  \providecommand\BibTeX{{%
    Bib\TeX}}}

\usepackage{amsmath,amssymb,amsfonts}
\usepackage{graphicx}
\usepackage{textcomp}
\usepackage{booktabs} 
\usepackage{multirow}
\usepackage{hyperref}
\usepackage{url}
\usepackage{nicefrac}
\usepackage{microtype}
\usepackage{xcolor}

\usepackage[utf8]{inputenc}
\usepackage{balance}
\usepackage{algorithm}
\usepackage{algpseudocode}
\usepackage{tikz}
\usetikzlibrary{positioning,arrows.meta,shapes.geometric,calc}
\usepackage{placeins}

\usepackage{orcidlink}
\usepackage{comment}
\usepackage{braket}
\usepackage{stfloats}
\def\BibTeX{{\rm B\kern-.05em{\sc i\kern-.025em b}\kern-.08em
    T\kern-.1667em\lower.7ex\hbox{E}\kern-.125emX}}

\usepackage{cite}
\usepackage{pifont}
\usepackage{amsmath,amssymb,amsfonts}
\usepackage{graphicx}
\usepackage{textcomp}
\usepackage{amsmath}
\usepackage{booktabs} 
\usepackage{color}
\usepackage{multirow}
\usepackage{colortbl, xcolor}
\usepackage{hyperref}       
\usepackage{url}            
\usepackage{booktabs}       
\usepackage{amsfonts}       
\usepackage{nicefrac}       
\usepackage{microtype}      
\usepackage{xcolor}         
\usepackage{algorithm}
\usepackage{algpseudocode}
\usepackage[font=small,skip=0pt]{caption}

\usepackage{amsmath}
\usepackage{tikz}
\usepackage{stfloats}
\usepackage{orcidlink}
\usepackage{comment}
\usepackage{amssymb}
\usepackage{braket}
\usepackage{xcolor}
\usepackage{cleveref}

\usepackage{enumitem}

\usepackage{setspace}
\begin{document}
\title{A2QTGN: Adaptive Amplitude Quantum-Integrated Temporal Graph Network  for Dynamic Link Prediction}

\author{\IEEEauthorblockN{Nouhaila Innan\orcidlink{0000-0002-1014-3457}\textsuperscript{1,2}, M. Murali	Karthick\orcidlink{0009-0000-6463-6065}\textsuperscript{3}, Simeon Kandan Sonar\orcidlink{0009-0002-0088-3547}\textsuperscript{3}, \\Vivek	Chaturvedi\orcidlink{0000-0003-1358-0107}\textsuperscript{3}, Muhammad Shafique\orcidlink{0000-0002-2607-8135}\textsuperscript{1,2}\\
 \IEEEauthorblockA{
\textsuperscript{1}eBRAIN Lab, Division of Engineering, New York University Abu Dhabi (NYUAD), Abu Dhabi, UAE\\
\textsuperscript{2}Center for Quantum and Topological Systems (CQTS), NYUAD Research Institute, NYUAD, Abu Dhabi, UAE\\
\textsuperscript{3}Indian Institute of Technology Palakkad (IITPKD), Palakkad 678623, India\\
\{nouhaila.innan, muhammad.shafique\}@nyu.edu, \{112301019,112301031\}@smail.iitpkd.ac.in, vivek@iitpkd.ac.in\\
}}}

\maketitle
\thispagestyle{empty}
\pagestyle{empty}
\begin{abstract}
Dynamic link prediction is important for modeling evolving interactions in social, communication, financial, and transportation networks. Classical temporal graph models capture changes over time, but they may struggle to represent rapidly evolving node-edge interactions in large dynamic graphs. We propose A2QTGN (Adaptive Amplitude Quantum-Integrated Temporal Graph Network), a hybrid quantum-classical framework that introduces adaptive amplitude encoding as a temporal embedding layer within a Temporal Graph Network. Unlike fixed quantum embeddings, the proposed module maps temporally varying node features into quantum states and selectively refreshes their amplitude representations according to the magnitude of feature change. This allows the framework to preserve stable node information, emphasize meaningful temporal variations, and reduce redundant quantum re-encoding. Across five Temporal Graph Benchmark datasets, A2QTGN achieves test area under the curve (AUC) values of up to 0.9957 and mean reciprocal rank (MRR) values of up to 0.7832, including the highest MRR among the compared baselines on tgbl-review and tgbl-flight. Ablation results further show that the adaptive quantum embedding is central to the model performance: on a 25k-event subset of tgbl-wiki, it improves test accuracy by 13.36 percentage points over always updating and by 22.44 percentage points over using no updates. The trained model is also evaluated using noisy simulations based on an IBM quantum device, followed by a smaller real-device experiment. The results show that adaptive quantum embeddings can provide effective temporal representations for dynamic link prediction while remaining executable under current quantum hardware constraints.
\end{abstract}




\section{Introduction}

\begin{figure}[htpb]
 \centering
 \includegraphics[height=2.8cm, width=9cm]{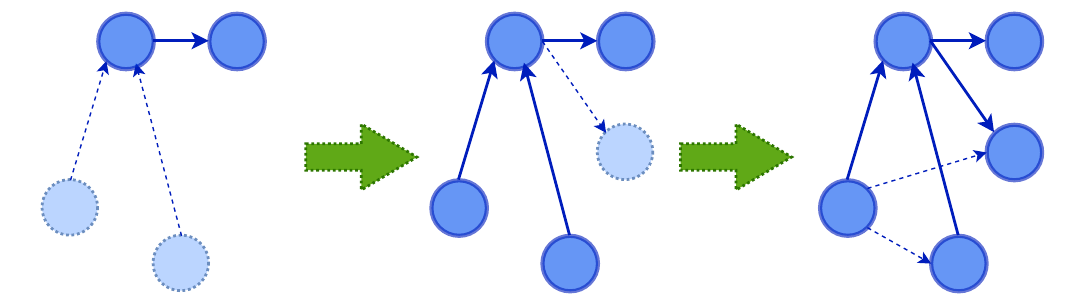}
 \caption{Dynamic graphs}
 \label{fig:dynamic_graph}
 \end{figure}

Dynamic graphs, which represent systems whose structure changes over time, capture sequences of graph snapshots where both nodes and edges may appear, disappear, or update their attributes \cite{vsiljak2008dynamic}. As illustrated in Fig.~\ref{fig:dynamic_graph}, such representations naturally arise in social platforms, recommender systems, financial transaction ecosystems, and large-scale transportation networks. A central computational challenge in these domains is \textit{dynamic link prediction} \cite{poursafaei2022towards}, the task of forecasting future edge formation or dissolution. Accurate prediction enables critical downstream applications, including friend recommendation, fraud detection, anomaly monitoring, and biological network analysis.
Conventional approaches, including static Graph Neural Network (GNN) \cite{9046288}, temporal GNNs, autoregressive sequence models, and heuristic predictors, often struggle to capture long-range temporal dependencies, structural drift, and the non-stationary behaviors characteristic of evolving relational systems \cite{kazemi2020representation,skarding2021foundations}. These limitations are further amplified by fairness and bias concerns, as undesirable patterns can accumulate or propagate across successive graph states, ultimately constraining reliable deployment in real-world settings.

Recent progress in quantum embedding techniques provides a promising avenue for addressing these gaps \cite{lloyd2020quantum,innan2026embeddings}. Instead of relying on claims of exponential advantage, quantum-inspired feature maps offer alternative geometric representations by transforming classical inputs into structured quantum states whose inner-product relationships may reveal patterns not easily captured by standard encoders. These embeddings modify the underlying decision boundary in feature space, potentially enabling models to distinguish subtle structural variations in evolving graphs.

Prior hybrid studies have demonstrated that augmenting classical architectures with parametrized quantum circuits can improve representation quality, introduce non-linear feature transformations, and stabilize optimization dynamics \cite{schuld2019quantum,havlivcek2019supervised,xu2024quantum,innan2024financial,innan2026spate}. These observations motivate deeper exploration into quantum-enhanced temporal graph models.

However, most existing quantum-graph approaches focus on static graphs and use standard methods, such as angle and amplitude encoding, without adapting the quantum representation as the graph changes. When applied to temporal graphs, these methods may re-encode node features at every time step, even when changes are small, leading to unnecessary quantum circuit executions. This motivates a temporally responsive quantum embedding that updates only when meaningful changes occur.

To address this limitation, we propose \textbf{A2QTGN}, an \textit{Adaptive Amplitude Quantum-Integrated Temporal Graph Network} that combines adaptive amplitude encoding with a classical Temporal Graph Network (TGN) backbone. A2QTGN measures changes in node features and refreshes the quantum representation only when the change exceeds an activity threshold. This preserves stable node states, reduces redundant quantum circuit executions, and integrates adaptive quantum embeddings with temporal memory propagation.

\textbf{Our main contributions are summarized as follows:}
\begin{enumerate}
    \item We propose \textbf{Adaptive Amplitude Encoding (AAE)}, a dynamic quantum embedding mechanism that updates amplitude representations according to temporal feature variations while avoiding unnecessary quantum re-encoding.

    \item We design a \textbf{hybrid Quantum--TGN pipeline} that combines AAE with a lightweight TGN backbone for end-to-end learning on evolving graph streams without increasing circuit depth with graph size.

    \item We introduce an \textbf{adaptive quantum update strategy} that determines when quantum embeddings should be refreshed, improving generalization over always-update and no-update variants.

    \item We provide a \textbf{comprehensive evaluation} on five TGBL link-prediction datasets, including baseline comparisons, ablation studies, convergence analysis, and hardware-aware inference with a limited real-device execution check.
\end{enumerate}


\section{Related Work \label{relatedwork}}

Prior studies on link prediction can be systematically categorized into three major methodological strands: those that capture time-evolving relationships through \textbf{temporal graph networks}, those employing established techniques such as \textbf{classical link predictors and meta-learning} approaches, and those exploring the potential of quantum computation with \textbf{quantum link prediction methods}.

\subsection{Temporal Graph Networks}

\noindent The evolution of temporal graph networks began with models like JODIE~\cite{10.1145/3292500.3330895}, which used recurrent neural networks to model continuous-time dynamics, but its sequential dependency incurred high computational cost and limited scalability. This led to the foundational TGN~\cite{rossi2020temporal}, which introduced an efficient, event-driven message passing mechanism coupled with a node memory module for inductive learning on streaming graphs. However, TGN's design suffered from two primary shortcomings which were limited expressive power due to simple aggregation, and poor generalization across diverse networks. Post-TGN advancements addressed these issues through architectural diversification: TGAT~\cite{xu2020inductiverepresentationlearningtemporal} utilized self-attention for better aggregation but faced quadratic complexity; PINT~\cite{souza2022provablyexpressivetemporalgraph} enforced provably injective message passing to enhance expressivity; and lightweight models like GraphMixer~\cite{cong2023reallyneedcomplicatedmodel} favored efficiency using MLPs but often sacrificed performance on complex dynamics. The latest trend, exemplified by research into  TGS~\cite{shamsi2025mintmultinetworktrainingtransfer}, aims to overcome the generalization barrier by using large-scale pre-training to establish transferable knowledge and neural scaling laws for dynamic graphs.

\subsection{Classical Link Predictors, Embeddings, and Meta-Learning}

Link prediction in static and dynamic graphs remains a challenge, as extensively studied by Ghasemian \textit{et al.}~\cite{ghasemian2020stacking}. Their evaluation across hundreds of real-world networks demonstrated that no single algorithm dominates, leading to the consensus that near-optimal performance relies on meta-learning ensembles combining multiple structural and embedding-based methods. While models that integrate node attributes, such as AGEE~\cite{Gu_2023}, successfully push performance yielding a 3\% improvement over conventional embedding methods, the struggle for efficiency and expressivity persists in increasingly complex scenarios. For instance, methods designed for highly expressive knowledge graph link prediction, such as ComplEx~\cite{trouillon2016complexembeddingssimplelink} or RotatE~\cite{sun2019rotateknowledgegraphembedding}, achieved high scores but are fundamentally constrained by their transductive nature and high computational demands, struggling to scale to the billions of edges found in massive industrial graphs. Similarly, while deep GNN architectures like \textbf{GraIL}~\cite{teru2020inductiverelationpredictionsubgraph} enable inductive link prediction, they require expensive local sub-graph extraction for every link prediction query, drastically increasing the inference latency and posing a clear, quantitative trade-off between structural expressivity and the necessity of real-time computational efficiency. This ongoing friction necessitates continued research into scalable yet highly expressive models.

\subsection{Quantum Link Prediction and Embeddings}

Recent efforts have explored quantum models for link prediction and representation learning. Fundamental to this area are quantum encoding schemes, including basic strategies such as angle encoding~\cite{tudisco2025evaluatingangleamplitudeencoding} and Amplitude encoding~\cite{tudisco2025evaluatingangleamplitudeencoding}, as well as more complex circuit-based methods, including Instantaneous Quantum Polynomial (IQP) circuits and Quantum Approximate Optimization Algorithm (QAOA)–inspired circuits, all of which map classical features into expressive quantum states.
Building upon these foundational encoding techniques, specialized quantum models have been developed for relational tasks. PQKELP~\cite{KUMAR2025125944}, for instance, introduced a projected quantum kernel embedding approach tailored for dynamic networks, demonstrating that quantum feature mappings can significantly improve relational reasoning capabilities. Concurrently, Moutinho \textit{et al.}~\cite{PhysRevA.107.032605} proposed a quantum-walk-based link prediction model, effectively showing how quantum interference effects can be harnessed to uncover latent connectivity patterns.

Despite these critical advances, existing quantum approaches predominantly operate on static graphs. They fundamentally lack dedicated mechanisms for temporal memory or continuous-time node representation, limiting their ability to effectively capture long-term dependencies in evolving graph structures.

To address these gaps, we introduce the \textbf{A2QTGN}, which integrates adaptive quantum encoding within a temporal graph framework to enable dynamic node-state modeling and efficient temporal information propagation.

\section{Methodology}
\label{methodology}

The proposed framework, \textbf{A2QTGN}, integrates adaptive quantum feature encoding within a temporal graph learning architecture for dynamic link prediction on evolving graphs. The model consists of two main components:
(1) the \textit{AAE} module, which converts temporal changes in node features into adaptive quantum embeddings; and
(2) the \textit{TGN}, which fuses these embeddings with node memory to model temporal dependencies and predict future links.

\begin{figure*}[htpb]
    \centering
    \includegraphics[width=1\linewidth]{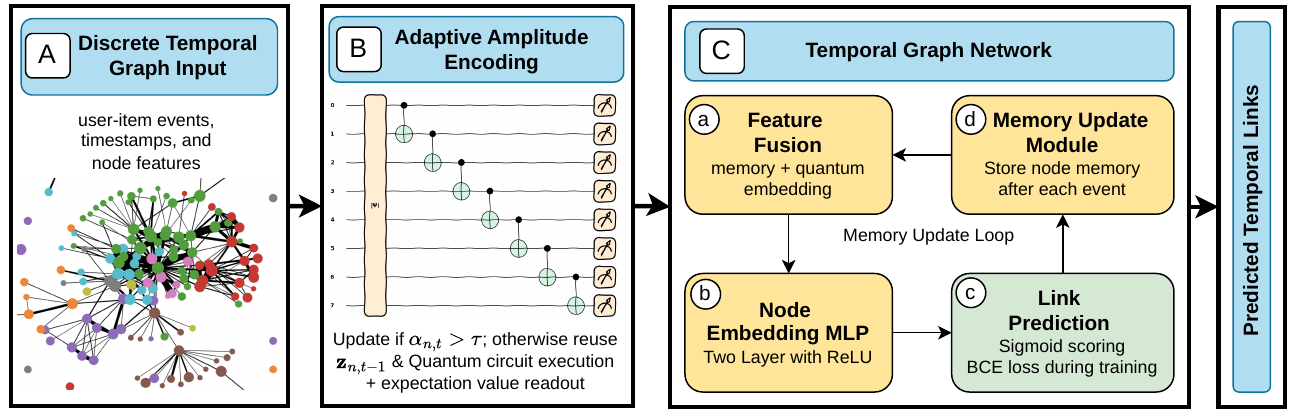}
    \caption{Overview of the A2QTGN pipeline. \textbf{(A)} Discrete temporal graph input containing user-item interactions, timestamps, and node features. \textbf{(B)} Adaptive Amplitude Encoding, which conditionally updates the quantum embedding according to temporal feature changes. \textbf{(C)} Temporal Graph Network, which performs feature fusion, node embedding, link prediction, and memory update to produce the predicted temporal links.}
    \label{fig:pipeline}
\end{figure*}

As shown in Fig.~\ref{fig:pipeline}, the workflow starts from a discrete temporal graph input, followed by adaptive amplitude encoding, and then a TGN stage that performs feature fusion, node embedding, link prediction, and memory update. The final output is the predicted temporal links. The following subsections describe each component in detail.

\subsection{Discrete Temporal Graph Formulation}

A2QTGN processes an ordered stream of temporal interactions represented as a sequence of graph states:
\begin{equation}
\mathcal{G}
=
\left\{
\mathcal{G}_t
=
\left(
\mathcal{V}_t,
\mathcal{E}_t,
\mathbf{X}_t
\right)
\right\}_{t=1}^{T},
\end{equation}
where $t$ denotes the current time step and $T$ is the total number of observed time steps. With $\mathcal{V}_t$ is the set of nodes, $\mathcal{E}_t\subseteq\mathcal{V}_t\times\mathcal{V}_t$ is the set of observed edges, and $\mathbf{X}_t$ is the node feature matrix at time $t$. Each interaction event $(u_t,i_t,t)$ represents an interaction between user $u_t$ and item $i_t$ at time $t$, along with their associated node features.

Given the interaction history
$\{\mathcal{G}_{s}\}_{s\leq t}$,
where $s$ indexes the observed time steps up to $t$, the dynamic link prediction objective is to estimate the probability that user $u$ will interact with item $i$ at a future time $t'>t$:
\begin{equation}
\hat{p}_{u,i,t'}
=
\mathbb{P}
\left(
(u,i)\in\mathcal{E}_{t'}
\mid
\{\mathcal{G}_{s}\}_{s\leq t}
\right),
\qquad
t'>t,
\end{equation}
where $\hat{p}_{u,i,t'}\in[0,1]$ denotes the predicted probability of the future interaction. This formulation defines the temporal graph input used by the Adaptive Amplitude Encoding and Temporal Graph Network components of A2QTGN.

\subsection{Adaptive Amplitude Encoding}

The AAE module is the main quantum contribution of A2QTGN. As shown in Fig.~\ref{fig:pipeline}, AAE is applied after the discrete temporal graph input and before the TGN. Its role is to transform temporal changes in node features into quantum embeddings that are later fused with the node memories maintained by the TGN.

Standard amplitude encoding requires node features to be encoded whenever a new graph state is processed. In temporal graphs, however, many node features remain unchanged or vary only slightly between consecutive interactions. AAE therefore measures the magnitude of each node's feature change and executes the quantum circuit only when the resulting activity factor exceeds a threshold. Otherwise, the previously computed quantum embedding is reused, reducing redundant circuit executions.

For a node $n$ at time $t$, let
$\mathbf{x}_{n,t}\in\mathbb{R}^{d}$
denote its feature vector, where $d$ is the classical feature dimension. The vector $\mathbf{x}_{n,t}$ corresponds to the row of the node feature matrix $\mathbf{X}_t$ associated with node $n$.

Let $t_n^{\mathrm{prev}}<t$ denote the most recent time at which node $n$ was processed. The temporal feature change is defined as
\begin{equation}
\Delta\mathbf{x}_{n,t}
=
\mathbf{x}_{n,t}
-
\mathbf{x}_{n,t_n^{\mathrm{prev}}}.
\end{equation}

The corresponding activity factor is
\begin{equation}
\alpha_{n,t}
=
\sigma
\left(
\beta
\left\|
\Delta\mathbf{x}_{n,t}
\right\|_2
\right),
\end{equation}
where $\sigma(s)=1/(1+e^{-s})$ is the sigmoid function, $\beta>0$ controls the sensitivity to feature changes, and $\tau\in(0,1)$ is the activity threshold.

When $\alpha_{n,t}>\tau$, the change-weighted feature vector is defined as
\begin{equation}
\mathbf{x}'_{n,t}
=
\frac{
\mathbf{x}_{n,t_n^{\mathrm{prev}}}
+
\alpha_{n,t}\Delta\mathbf{x}_{n,t}
}{
\left\|
\mathbf{x}_{n,t_n^{\mathrm{prev}}}
+
\alpha_{n,t}\Delta\mathbf{x}_{n,t}
\right\|_2
}.
\end{equation}
For a node with no previous embedding, the normalized current feature vector
$\mathbf{x}_{n,t}/\|\mathbf{x}_{n,t}\|_2$
is encoded directly.

Let $\mathcal{A}_t\subseteq\mathcal{V}_t$ denote the set of nodes whose embeddings are processed at time $t$. During training, this set includes nodes involved in observed interactions and nodes selected through negative sampling.

Algorithm~\ref{alg:aae} describes the proposed AAE module. The complete A2QTGN workflow, including temporal memory, link prediction, memory updating, and training, is presented in Algorithm~\ref{alg:a2qtgn}.

\begin{algorithm}[htpb]
\caption{Adaptive Amplitude Encoding}
\label{alg:aae}
\footnotesize
\begin{algorithmic}[1]

\Require Processed-node sets $\{\mathcal{A}_t\}_{t=1}^{T}$,
node features $\mathbf{x}_{n,t}\in\mathbb{R}^{d}$,
parameters $\beta$ and $\tau$,
and number of qubits $N_q$

\Ensure Quantum embeddings
$\{\mathbf{z}_{n,t}\in\mathbb{R}^{N_q}\}$

\For{each time step $t=1$ to $T$}
    \For{each node $n\in\mathcal{A}_t$}

        \If{node $n$ has no previous embedding}
            \State Normalize the current node features
            $\mathbf{x}_{n,t}$

            \State Encode the normalized features and execute the AAE circuit

            \State Measure the Pauli-$Z$ expectation values to obtain
            $\mathbf{z}_{n,t}$
        \Else
            \State Compute the temporal feature change
            $\Delta\mathbf{x}_{n,t}$

            \State Compute the activity factor
            $\alpha_{n,t}$

            \If{$\alpha_{n,t}>\tau$}
                \State Compute and normalize the change-weighted feature vector
                $\mathbf{x}'_{n,t}$

                \State Encode $\mathbf{x}'_{n,t}$ and execute the AAE circuit

                \State Measure the Pauli-$Z$ expectation values to obtain
                $\mathbf{z}_{n,t}$
            \Else
                \State Reuse the quantum embedding from the previous processing time
                $\mathbf{z}_{n,t_n^{\mathrm{prev}}}$
            \EndIf
        \EndIf

    \EndFor
\EndFor

\State \Return $\{\mathbf{z}_{n,t}\}$

\end{algorithmic}
\end{algorithm}

The operator $U_{\mathrm{AAE}}(\cdot)$ denotes the implemented AAE circuit. It maps a normalized input vector to the quantum state
$\lvert\psi_{n,t}\rangle$
using \texttt{AmplitudeEmbedding}~\cite{bergholm2018pennylane}, followed by controlled-NOT gates that introduce entanglement.

The expectation values of the Pauli-$Z$ operators form the quantum embedding:
\begin{equation}
\mathbf{z}_{n,t}
=
\left[
\langle\psi_{n,t}\rvert
\hat{Z}_1
\lvert\psi_{n,t}\rangle,
\ldots,
\langle\psi_{n,t}\rvert
\hat{Z}_{N_q}
\lvert\psi_{n,t}\rangle
\right]
\in
\mathbb{R}^{N_q},
\end{equation}
where $\hat{Z}_k$ is the Pauli-$Z$ operator measured on qubit $k$, and $N_q$ is the number of qubits and measured quantum features. When $\alpha_{n,t}>\tau$, the embedding reflects the change-weighted node features. Otherwise, the embedding from the previous processing time is retained. The resulting embedding $\mathbf{z}_{n,t}$ is passed to the TGN and fused with the node's previous memory.

\subsection{Temporal Graph Network}

The TGN component maintains the interaction history of each node through an event-driven memory mechanism. Each node has a memory vector that is retrieved when the node participates in an interaction and updated after link prediction. A2QTGN retains this memory structure from the classical TGN but modifies the node representation stage by integrating the adaptive quantum embedding generated by AAE. For a newly observed node, its memory is initialized as $\mathbf{m}_{n,0}=\mathbf{0}$.

As shown in Fig.~\ref{fig:pipeline}, the TGN component contains four stages.

\paragraph{Feature Fusion}

Let $n$ denote a node involved in an interaction at time $t$. Before processing the interaction, the node has a previous memory vector
$\mathbf{m}_{n,t}^{\mathrm{prev}}\in\mathbb{R}^{D_{\mathrm{mem}}}$,
where $D_{\mathrm{mem}}$ is the memory dimension. The AAE module provides the quantum embedding
$\mathbf{z}_{n,t}\in\mathbb{R}^{N_q}$,
where $N_q$ is the number of measured quantum features.

The previous memory and quantum embedding are concatenated to form
\begin{equation}
\mathbf{r}_{n,t}
=
\left[
\mathbf{m}_{n,t}^{\mathrm{prev}}
\parallel
\mathbf{z}_{n,t}
\right]
\in
\mathbb{R}^{D_{\mathrm{mem}}+N_q},
\end{equation}
where $\parallel$ denotes vector concatenation. AAE generates $\mathbf{z}_{n,t}$ from temporal feature changes, while the feature-fusion stage combines it with the node's stored interaction history.

\paragraph{Node Embedding MLP}

The fused representation $\mathbf{r}_{n,t}$ is passed through a two-layer multilayer perceptron with a ReLU activation:
\begin{equation}
\mathbf{h}_{n,t}
=
\mathrm{MLP}_{\mathrm{emb}}
\left(
\mathbf{r}_{n,t}
\right)
\in
\mathbb{R}^{D_h},
\end{equation}
where $\mathrm{MLP}_{\mathrm{emb}}$ denotes the node-embedding network and $D_h$ is the node-embedding dimension. We set $D_h=D_{\mathrm{mem}}$, allowing $\mathbf{h}_{n,t}$ to be stored directly as the updated node memory.

The complete embedding operation used in Algorithm~\ref{alg:a2qtgn} is therefore
\begin{equation}
\begin{aligned}
\mathbf{h}_{n,t}
&=
\mathrm{EmbeddingModule}
\left(
\mathbf{m}_{n,t}^{\mathrm{prev}},
\mathbf{z}_{n,t}
\right)
\\
&=
\mathrm{MLP}_{\mathrm{emb}}
\left(
\left[
\mathbf{m}_{n,t}^{\mathrm{prev}}
\parallel
\mathbf{z}_{n,t}
\right]
\right).
\end{aligned}
\end{equation}

The resulting representation $\mathbf{h}_{n,t}$ combines the node's historical memory with its current adaptive quantum embedding.

\paragraph{Link Prediction}

For a candidate interaction between user $u$ and item $i$ at time $t$, their embeddings $\mathbf{h}_{u,t}$ and $\mathbf{h}_{i,t}$ are concatenated and passed through a scoring network:
\begin{equation}
\hat{p}_{u,i,t}
=
\sigma
\left(
\mathrm{Scorer}
\left(
\left[
\mathbf{h}_{u,t}
\parallel
\mathbf{h}_{i,t}
\right]
\right)
\right),
\end{equation}
where $\mathrm{Scorer}$ is a multilayer perceptron, $\sigma(\cdot)$ is the sigmoid function, and $\hat{p}_{u,i,t}\in[0,1]$ is the predicted link probability.

During training, each prediction is optimized using binary cross-entropy:
\begin{equation}
\mathcal{L}_{u,i,t}
=
-y_{u,i,t}\log\hat{p}_{u,i,t}
-
\left(1-y_{u,i,t}\right)
\log\left(1-\hat{p}_{u,i,t}\right),
\end{equation}
where $y_{u,i,t}\in\{0,1\}$ is the ground-truth link label. During inference, $\hat{p}_{u,i,t}$ is used as the predicted probability of a future interaction.

\paragraph{Memory Update Module}

After link prediction, the embedding of each node involved in an observed interaction is stored as its updated memory: $\mathbf{m}_{n,t}
\leftarrow
\operatorname{stopgrad}
\left(
\mathbf{h}_{n,t}
\right),
\qquad
t_n^{\mathrm{last}}
\leftarrow
t,$
where $\mathbf{m}_{n,t}$ is the updated node memory, $t_n^{\mathrm{last}}$ records its most recent memory-update time, and $\operatorname{stopgrad}(\cdot)$ prevents gradients from propagating through earlier interactions. The updated memory is retrieved when node $n$ participates in a later interaction.

For each observed interaction $(u,i)$, a negative item
$i^{-}\in\mathcal{V}_t$ is sampled such that $(u,i^{-})\notin\bigcup_{s\leq t}\mathcal{E}_s.$
The corresponding positive and negative link probabilities are denoted by
$\hat{p}_{u,i,t}^{+}$ and $\hat{p}_{u,i^{-},t}^{-}$, respectively. The time-step loss $\mathcal{L}_t$ is computed from the binary cross-entropy losses of these positive and negative predictions.

Algorithm~\ref{alg:a2qtgn} summarizes the complete A2QTGN workflow, including adaptive quantum encoding, memory retrieval, node embedding formation, link prediction, loss computation, and memory updating.

\begin{algorithm}[htpb]
\caption{End-to-End A2QTGN Workflow}
\label{alg:a2qtgn}
\footnotesize
\begin{algorithmic}[1]

\Require Temporal graph sequence
$\{\mathcal{G}_t=
(\mathcal{V}_t,\mathcal{E}_t,\mathbf{X}_t)\}_{t=1}^{T}$,
memory modules, and AAE from Algorithm~\ref{alg:aae}

\Ensure Predicted link probabilities
$\{\hat{p}_{u,i,t}\}$

\State Initialize each node memory as
$\mathbf{m}_{n,0}=\mathbf{0}$

\For{each time step $t=1$ to $T$}
    \Comment{Process temporal interactions sequentially}

    \For{each observed interaction $(u,i)\in\mathcal{E}_t$}
        \State Sample a negative item $i^{-}$ for user $u$
    \EndFor

    \Statex \textbf{Node Processing}

    \For{each node $n$ involved in an observed or sampled interaction}
        \State Compute or reuse the quantum embedding
        $\mathbf{z}_{n,t}$ using Algorithm~\ref{alg:aae}

        \State Retrieve the previous node memory
        $\mathbf{m}_{n,t}^{\mathrm{prev}}$

        \State Fuse the previous memory and quantum embedding using the
        \texttt{EmbeddingModule}

        \State Obtain the temporal node embedding
        $\mathbf{h}_{n,t}$
    \EndFor

    \Statex \textbf{Link Prediction}

    \For{each observed user-item interaction $(u,i)\in\mathcal{E}_t$}
        \State Compute the positive link probability
        $\hat{p}_{u,i,t}^{+}$

        \State Compute the corresponding negative link probability
        $\hat{p}_{u,i^{-},t}^{-}$
    \EndFor

    \State Compute the binary cross-entropy loss
    $\mathcal{L}_t$ from the positive and negative link predictions

    \Statex \textbf{Memory Update}

    \For{each node $n$ involved in an observed interaction at time $t$}
        \State Update the node memory using the detached embedding
        $\mathbf{h}_{n,t}$

        \State Record $t$ as the node's latest memory-update time
    \EndFor

\EndFor

\State \Return $\{\hat{p}_{u,i,t}\}$

\end{algorithmic}
\end{algorithm}

\section{Results}
\label{results}

\subsection{Experimental Setup}

We evaluate A2QTGN on five temporal link-prediction datasets from the Temporal Graph Benchmark (TGB): \texttt{tgbl-wiki}, \texttt{tgbl-review}, \texttt{tgbl-coin}, \texttt{tgbl-comment}, and \texttt{tgbl-flight}~\cite{huang2024tgb2}. Their approximate statistics are reported in Table~\ref{tab:tgb_stats}.

\begin{table}[htpb]
\centering
\caption{Statistics of the TGBL datasets used for evaluation.}
\label{tab:tgb_stats}
\begin{tabular}{lcc}
\hline
\textbf{Dataset} & \textbf{\# Nodes} & \textbf{\# Edges} \\
\hline
\texttt{tgbl-wiki}    & 9,227   & 157,474 \\
\texttt{tgbl-review}  & 352,637 & 4,873,540 \\
\texttt{tgbl-coin}    & 638,486 & 22,809,486 \\
\texttt{tgbl-comment} & 994,790 & 44,314,507 \\
\texttt{tgbl-flight}  & 18,143  & 67,169,570 \\
\hline
\end{tabular}
\end{table}

The interactions are ordered chronologically and divided into 70\% for training, 15\% for validation, and 15\% for testing. During training, each positive interaction is paired with a sampled negative destination node for binary cross-entropy optimization. For validation and testing, we use the negative candidates provided by TGB to preserve a consistent evaluation setting across all models.
The model uses eight qubits, a 64-dimensional node memory, and a 64-dimensional node embedding. It is trained using the Adam optimizer with a learning rate of $10^{-3}$ and a batch size of 32.
The main experiments are conducted using the PennyLane \texttt{lightning.qubit} simulator. To evaluate performance under device noise, the trained model is tested on the \texttt{FakeTorino} noisy backend using 2048 shots. A separate smaller-scale experiment is executed on the real IBM \texttt{ibm\_torino} quantum device using 100 shots.

Evaluation is based on Mean Reciprocal Rank (MRR), Area under the curve (AUC), accuracy, and precision. \textbf{MRR} measures the rank of the true destination node among the candidate links and is computed as
\begin{equation}
\mathrm{MRR}
=
\frac{1}{|\mathcal{Q}|}
\sum_{q=1}^{|\mathcal{Q}|}
\frac{1}{\mathrm{rank}_q},
\end{equation}
where $\mathcal{Q}$ is the set of evaluation queries and $\mathrm{rank}_q$ is the rank of the true destination node for query $q$.

\textbf{AUC} evaluates the separation between observed and negative links, while \textbf{accuracy} measures the proportion of correctly classified links. \textbf{precision} indicates the reliability of predicted positive links by reflecting the effect of false-positive predictions. These metrics assess both candidate ranking and binary link-prediction quality.
\subsection{Performance Analysis}
The objective of this experiment is to evaluate how effectively A2QTGN models temporal interactions and predicts future links across diverse real-world dynamic graphs. Table~\ref{tab:tgb_main_results} reports the train, validation, and test performance of A2QTGN on the five TGBL datasets using accuracy, precision, AUC, and MRR.
\begin{table}[h]
\centering
\caption{Train, validation, and test performance of A2QTGN on TGBL datasets.}
\label{tab:tgb_main_results}
\resizebox{\linewidth}{!}{%
\begin{tabular}{llccccc}
\toprule
\textbf{Split} & \textbf{Metric} & \textbf{Wiki} & \textbf{Review} & \textbf{Coin} & \textbf{Comment} & \textbf{Flight} \\
\midrule

\multirow{4}{*}{Train}
& Accuracy  & 92.72\% & 71.13\% & 73.66\% & 81.40\% & 96.67\% \\
& Precision & 90.49\% & 76.81\% & 85.25\% & 82.81\% & 95.31\% \\
& AUC       & 0.9809  & 0.7665  & 0.8050  & 0.8842  & 0.9921 \\
& MRR       & 0.7651  & 0.5732  & 0.6125  & 0.6642  & 0.7811 \\
\midrule

\multirow{4}{*}{Validation}
& Accuracy  & 92.99\% & 68.69\% & 78.61\% & 80.63\% & 97.35\% \\
& Precision & 90.25\% & 69.41\% & 80.21\% & 77.84\% & 95.33\% \\
& AUC       & 0.9845  & 0.7921  & 0.8784  & 0.8817  & 0.9922 \\
& MRR       & 0.7674  & 0.6142  & 0.6624  & 0.6409  & 0.7808 \\
\midrule

\multirow{4}{*}{Test}
& Accuracy  & 93.03\% & 64.99\% & 79.87\% & 79.67\% & 97.83\% \\
& Precision & 90.13\% & 65.76\% & 80.97\% & 77.48\% & 96.07\% \\
& AUC       & 0.9857  & 0.7657  & 0.8886  & 0.8725  & 0.9957 \\
& MRR       & 0.7782  & 0.5883  & 0.6668  & 0.6383  & 0.7832 \\
\bottomrule
\end{tabular}}
\end{table}
A2QTGN achieves its highest test performance on \texttt{tgbl-flight}, with 97.83\% accuracy, 96.07\% precision, an AUC of 0.9957, and an MRR of 0.7832. Strong results are also obtained on \texttt{tgbl-wiki}, where the model reaches an AUC of 0.9857 and an MRR of 0.7782. In contrast, \texttt{tgbl-review} presents the lowest test performance, with an AUC of 0.7657 and an MRR of 0.5883, showing that the prediction difficulty varies across datasets.

The results remain relatively stable across the train, validation, and test splits for \texttt{tgbl-wiki}, \texttt{tgbl-comment}, and \texttt{tgbl-flight}. The largest reduction from training to testing occurs on \texttt{tgbl-review}, particularly in accuracy and precision. For \texttt{tgbl-coin}, the AUC increases from 0.8050 during training to 0.8886 during testing, while the MRR increases from 0.6125 to 0.6668. These results show that A2QTGN maintains effective ranking and link-discrimination performance across later temporal periods.

\subsection{Algorithmic Convergence Performance}

\begin{figure*}[htpb]
    \centering
    \includegraphics[width=1\linewidth]{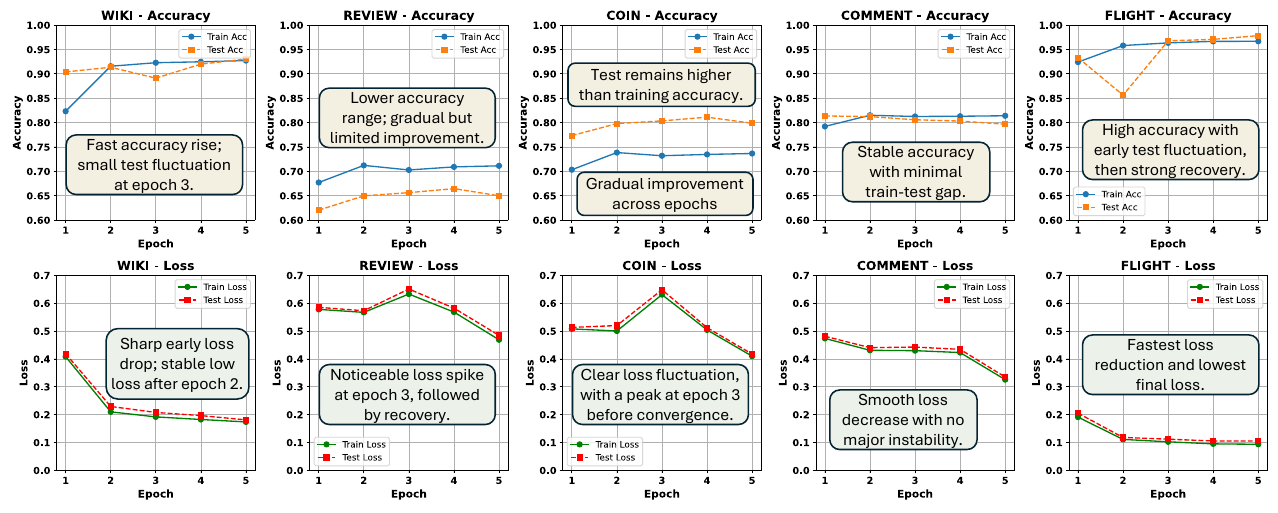}
    \caption{Training and validation convergence of A2QTGN across the evaluated datasets.}
    \label{fig:res_plots}
\end{figure*}

This experiment evaluates the convergence speed and training stability of A2QTGN across temporal graphs of different scales. As shown in Fig.~\ref{fig:res_plots}, the model converges rapidly on \texttt{tgbl-wiki} and maintains smooth learning behavior on larger datasets such as \texttt{tgbl-coin}. The training and validation curves remain close across epochs, indicating limited overfitting under the considered settings.

The stable convergence is also consistent with the selective update mechanism of A2QTGN, which refreshes the quantum embedding only when the temporal feature change exceeds the activity threshold. This reduces redundant quantum evaluations while preserving the feature updates needed for model optimization.
\subsection{Ablation Study}

The ablation study examines the contributions of the quantum embedding and the adaptive update mechanism. The proposed A2QTGN model refreshes the quantum embedding only when the activity factor exceeds the update threshold. \textit{Classical TGN} removes the quantum embedding and relies only on temporal memory, while \textit{Always Update} retains the quantum module but recomputes the embedding at every interaction. Additional experiments include a \textit{No Update} variant, which computes the initial quantum embedding once and reuses it throughout the temporal sequence. These configurations separate the effect of quantum feature encoding from the effect of conditionally updating the resulting representation.

\begin{table}[htbp]
\centering
\caption{Main ablation results on the \texttt{tgbl-wiki} dataset.}
\label{tab:ablation_results}
\resizebox{\columnwidth}{!}{%
\begin{tabular}{lcccc}
\hline
\textbf{Model Variant} & \textbf{Accuracy} & \textbf{Precision} & \textbf{AUC} & \textbf{MRR} \\
\hline
A2QTGN      & 93.03\% & 90.13\% & 0.9857 & 0.7782 \\
Classical TGN & 54.12\% & 52.91\% & 0.6014 & 0.2187 \\
Always Update & 68.47\% & 67.20\% & 0.7258 & 0.3421 \\
\hline
\end{tabular}}
\end{table}

As reported in Table~\ref{tab:ablation_results}, A2QTGN achieves the highest performance across all metrics. Removing the quantum embedding produces the largest reduction, with Classical TGN decreasing the AUC from 0.9857 to 0.6014 and the MRR from 0.7782 to 0.2187. Under the same temporal learning setting, this result shows that temporal memory alone does not provide the representation quality achieved when the adaptive quantum embedding is included.

Always update also performs below A2QTGN despite retaining the quantum module. Its AUC decreases to 0.7258 and its MRR to 0.3421. This comparison shows that the improvement does not arise only from quantum encoding, but also from controlling when the embedding is refreshed. Updating at every interaction treats small and substantial feature changes in the same manner, which may repeatedly alter the representation even when a new quantum evaluation is unnecessary.

To examine the update mechanism at different data scales, additional experiments are conducted on 10k and 25k subsets of \texttt{tgbl-wiki}. These experiments compare A2QTGN with the always update and no update variants.

\begin{table}[htbp]
\centering
\caption{Update-strategy ablation on the 10k subset of the \texttt{tgbl-wiki} dataset.}
\label{tab:ablation_10k}
\resizebox{\columnwidth}{!}{%
\begin{tabular}{llcccc}
\hline
\textbf{Split} & \textbf{Model Variant} & \textbf{Accuracy} & \textbf{Precision} & \textbf{AUC} & \textbf{MRR} \\
\hline
\multirow{3}{*}{Train}
& A2QTGN       & 91.71\% & 88.74\% & 0.9759 & 0.7583 \\
& Always Update & 83.95\% & 81.56\% & 0.9166 & 0.6912 \\
& No Update     & 83.51\% & 82.57\% & 0.9134 & 0.6943 \\
\hline
\multirow{3}{*}{Val}
& A2QTGN       & 92.07\% & 89.03\% & 0.9773 & 0.7678 \\
& Always Update & 81.50\% & 83.99\% & 0.9067 & 0.6915 \\
& No Update     & 70.13\% & 79.94\% & 0.7909 & 0.6809 \\
\hline
\multirow{3}{*}{Test}
& A2QTGN       & 92.52\% & 89.17\% & 0.9798 & 0.7782 \\
& Always Update & 80.28\% & 83.80\% & 0.8955 & 0.6877 \\
& No Update     & 68.79\% & 80.63\% & 0.7528 & 0.6351 \\
\hline
\end{tabular}}
\end{table}

\begin{table}[htbp]
\centering
\caption{Update-strategy ablation on the 25k subset of the \texttt{tgbl-wiki} dataset.}
\label{tab:ablation_25k}
\resizebox{\columnwidth}{!}{%
\begin{tabular}{llcccc}
\hline
\textbf{Split} & \textbf{Model Variant} & \textbf{Accuracy} & \textbf{Precision} & \textbf{AUC} & \textbf{MRR} \\
\hline
\multirow{3}{*}{Train}
& A2QTGN       & 91.21\% & 88.56\% & 0.9751 & 0.7704 \\
& Always Update & 81.43\% & 80.82\% & 0.9012 & 0.6735 \\
& No Update     & 86.72\% & 87.71\% & 0.9391 & 0.7282 \\
\hline
\multirow{3}{*}{Val}
& A2QTGN       & 92.09\% & 90.44\% & 0.9783 & 0.7693 \\
& Always Update & 78.61\% & 76.39\% & 0.8865 & 0.6671 \\
& No Update     & 67.53\% & 78.65\% & 0.7767 & 0.6231 \\
\hline
\multirow{3}{*}{Test}
& A2QTGN       & 90.63\% & 89.05\% & 0.9739 & 0.7597 \\
& Always Update & 77.27\% & 75.35\% & 0.8606 & 0.6458 \\
& No Update     & 68.19\% & 74.60\% & 0.7479 & 0.6203 \\
\hline
\end{tabular}}
\end{table}

A2QTGN achieves the highest validation and test performance on both subsets. On the 25k test split, it reaches 90.63\% accuracy, 0.9739 AUC, and 0.7597 MRR. Compared with always update, A2QTGN improves accuracy by 13.36 percentage points and MRR by 0.1139. Compared with no update, the corresponding gains are 22.44 percentage points and 0.1394.

The results expose the limitations of both fixed update strategies. Always update refreshes the quantum embedding even when temporal feature changes are limited, which can reduce representation stability and increase redundant circuit evaluations. No update preserves a fixed embedding that cannot reflect later changes in node behavior, causing the representation to become outdated as the graph evolves. A2QTGN balances stability and temporal responsiveness by retaining the previous embedding when changes are limited and recomputing it when the activity threshold is exceeded.
\subsection{Comparative Analysis}

We conduct a two-level comparison to evaluate A2QTGN against alternative quantum embedding techniques and representative temporal graph learning models.

\begin{table}[htbp]
\centering
\caption{Performance comparison of quantum embedding techniques on the \texttt{tgbl-wiki} dataset.}
\label{tab:embedding_comparison}
\resizebox{\columnwidth}{!}{%
\begin{tabular}{lcccc}
\hline
\textbf{Model} & \textbf{Accuracy} & \textbf{Precision} & \textbf{AUC} & \textbf{MRR} \\
\hline
Angle Embedding + TGN     & 65.91\% & 39.59\% & 0.7086 & 0.3438 \\
Amplitude Embedding + TGN & 74.96\% & 46.00\% & 0.8056 & 0.3199 \\
IQP Embedding \cite{havlivcek2019supervised} + TGN       & 49.73\% & 23.67\% & 0.5349 & 0.1638 \\
QAOA Embedding \cite{lloyd2020quantum} + TGN      & 68.04\% & 34.68\% & 0.7313 & 0.2828 \\
\textbf{A2QTGN (Ours)}    & \textbf{93.03\%} & \textbf{90.13\%} &
\textbf{0.9857} & \textbf{0.7782} \\
\hline
\end{tabular}}
\end{table}

\paragraph{Quantum-Embedding Comparison}

We first examine how the choice of quantum embedding affects temporal link prediction while keeping the same TGN backbone, data split, training settings, and evaluation protocol. The compared methods differ in how node features are represented in the quantum circuit, as reported in Table~\ref{tab:embedding_comparison}.
\begin{table*}[b]
\centering
\caption{MRR comparison across the complete TGBL datasets.}
\label{tab:mrr_tgbl}
\resizebox{\linewidth}{!}{%
\begin{tabular}{l *{9}{c}}
\toprule
\textbf{Dataset} &
EdgeBank (tw)~\cite{yu2023towards} &
EdgeBank (un)~\cite{yu2023towards} &
DyRep~\cite{trivedi2019dyrep} &
TGN~\cite{rossi2020temporal} &
CAWN~\cite{wang2021inductive} &
NAT~\cite{luo2022neighborhood} &
DyGFormer~\cite{yu2023towards} &
TGAT~\cite{xu2020inductive} &
\textbf{A2QTGN} \\
\midrule
\texttt{tgbl-wiki}
& 0.571 & 0.495 & 0.050 & 0.396 & 0.711 & 0.749 &
\textbf{0.798} & 0.141 & 0.7782 \\

\texttt{tgbl-review}
& 0.025 & 0.023 & 0.220 & 0.349 & 0.193 & 0.341 &
0.224 & 0.355 & \textbf{0.5883} \\

\texttt{tgbl-coin}
& 0.580 & 0.359 & 0.452 & 0.586 & -- & -- &
\textbf{0.752} & -- & 0.6688 \\

\texttt{tgbl-comment}
& 0.149 & 0.129 & 0.289 & 0.379 & -- & -- &
\textbf{0.670} & -- & 0.6383 \\

\texttt{tgbl-flight}
& 0.387 & 0.167 & 0.556 & 0.705 & -- & -- &
-- & -- & \textbf{0.7832} \\
\bottomrule
\end{tabular}}
\end{table*}
Among the fixed quantum embeddings, \textbf{Amplitude embedding} achieves the highest accuracy and AUC, suggesting that its compact global representation is more suitable for the evaluated node features. \textbf{Angle embedding} represents features mainly through separate local rotations, which may limit its ability to capture relationships across the complete feature vector. \textbf{IQP embedding} introduces phase-based feature interactions, but these interactions do not produce sufficiently separable measurement outputs under the evaluated setting. \textbf{QAOA embedding} adds a more complex trainable circuit structure, yet this additional optimization does not result in stronger temporal link representations. These observations are specific to the considered task and do not indicate that the embedding techniques perform poorly in other applications.

A2QTGN achieves the highest performance across all metrics. Unlike the fixed embedding baselines, it combines amplitude encoding with an adaptive decision rule that refreshes the quantum representation only when the temporal feature change exceeds the activity threshold. The results therefore indicate that the improvement arises from both the selected quantum representation and its conditional update mechanism.

\paragraph{Model-Level Comparison}

We next compare A2QTGN with representative temporal graph learning models on the TGBL datasets. The MRR results for EdgeBank, DyRep, TGN, CAWN, NAT, DyGFormer, TGAT, and A2QTGN are reported in Table~\ref{tab:mrr_tgbl}.

The evaluated models represent different approaches to temporal graph learning. EdgeBank predicts links from previously observed interactions. DyRep applies event-driven recurrent updates to node representations, while TGN maintains explicit node memories over temporal events. TGAT uses temporal attention over historical neighborhoods, CAWN models temporal structure through causal anonymous walks, and NAT uses neighborhood-aware representations. DyGFormer combines neighbor co-occurrence encoding with sequence patching to model long interaction histories.

The symbol ``--'' indicates that no result was reported for the corresponding model-dataset pair in the cited benchmark source. The original benchmark attributes several missing results on the larger datasets to memory or computational limitations.

A2QTGN achieves the highest MRR on \texttt{tgbl-review} and \texttt{tgbl-flight}. On \texttt{tgbl-wiki}, it reaches 0.7782, close to DyGFormer at 0.798 and higher than the remaining baselines. On \texttt{tgbl-coin}, A2QTGN obtains 0.6688, outperforming EdgeBank, DyRep, and TGN, although DyGFormer reaches 0.752. On \texttt{tgbl-comment}, its MRR of 0.6383 is close to DyGFormer at 0.670 and exceeds all other available results. These findings show that A2QTGN maintains strong ranking performance across datasets with different temporal patterns and interaction volumes.

\subsection{Noisy-Backend and Real-Hardware Inference Evaluation}

We evaluate A2QTGN using the model trained on the 25k-interaction setting under both noisy simulation and physical quantum hardware execution.

The noisy evaluation uses \texttt{FakeTorino} with 2048 shots per circuit. \texttt{FakeTorino} is a Qiskit simulator configured from stored properties of the \textbf{IBM} Torino device, including its qubit connectivity, supported gates, and calibration-based gate and measurement noise. This provides repeatable device-informed evaluation without executing on a physical processor. The metrics are computed over 100 temporal link samples.

We also execute the trained quantum circuits on the physical IBM \texttt{ibm\_torino} processor using 100 shots per circuit.

\begin{table}[htbp]
\centering
\caption{Inference performance on \texttt{FakeTorino} using the 25k-interaction setting, 2048 shots per circuit, and 100 evaluation samples.}
\label{tab:torino_results}
\resizebox{\columnwidth}{!}{%
\begin{tabular}{lccccc}
\toprule
\textbf{Metric} & \textbf{Wiki} & \textbf{Review} & \textbf{Coin} &
\textbf{Comment} & \textbf{Flight} \\
\midrule
Accuracy  & 86.50\% & 67.00\% & 77.00\% & 73.50\% & 80.00\% \\
Precision & 91.95\% & 77.00\% & 73.68\% & 70.00\% & 94.12\% \\
AUC       & 0.9631  & 0.7676  & 0.8586  & 0.7222  & 0.9654 \\
MRR       & 0.9778  & 0.8895  & 0.9154  & 0.8226  & 0.9844 \\
\bottomrule
\end{tabular}}
\end{table}

A2QTGN retains strong performance under the \texttt{FakeTorino} noise model. It achieves an AUC of 0.9631 and an MRR of 0.9778 on \texttt{tgbl-wiki}, while reaching an AUC of 0.9654 and an MRR of 0.9844 on \texttt{tgbl-flight}. The model also obtains an MRR of 0.9154 on \texttt{tgbl-coin}. Lower accuracy and AUC values on \texttt{tgbl-review} and \texttt{tgbl-comment} indicate that sensitivity to finite-shot and backend noise varies across datasets.

On the physical \texttt{ibm\_torino} processor, A2QTGN achieves 90.00\% accuracy and an AUC of 1.00 on \texttt{tgbl-coin}, and 70.00\% accuracy with an AUC of 0.80 on \texttt{tgbl-review}. These results verify the execution of the trained quantum inference circuits on a real IBM quantum processor.
\section{Discussion and Conclusion}
\label{sec:discussion_conclusion}

A2QTGN achieves strong temporal link-prediction performance across datasets with different sizes and interaction patterns. It reaches an AUC of 0.9957 and an MRR of 0.7832 on \texttt{tgbl-flight}, together with an AUC of 0.9857 and an MRR of 0.7782 on \texttt{tgbl-wiki}. In contrast, \texttt{tgbl-review} remains more challenging, with an AUC of 0.7657 and an MRR of 0.5883. Since \texttt{tgbl-flight} contains the largest number of interactions but achieves the strongest results, dataset difficulty depends not only on graph size but also on temporal regularity and link separability.

The reported metrics capture different prediction objectives. Accuracy, precision, and AUC evaluate binary link discrimination, whereas MRR measures the rank of the true destination among candidate nodes. This distinction explains why the \texttt{FakeTorino} evaluation retains high MRR values, such as 0.9778 on \texttt{tgbl-wiki} and 0.9844 on \texttt{tgbl-flight}, despite lower accuracy values. These results are computed over 100 samples from the 25k-interaction setting and are therefore not directly comparable with the complete-dataset evaluation.

The ablation study confirms the importance of adaptive quantum updates. On the 25k-interaction \texttt{tgbl-wiki} setting, A2QTGN achieves 90.63\% accuracy and an MRR of 0.7597, compared with 77.27\% and 0.6458 for always updating, and 68.19\% and 0.6203 when the embedding is never updated. Selective updating therefore balances temporal responsiveness with reduced circuit execution.

The \texttt{FakeTorino} and \texttt{ibm\_torino} experiments further verify execution under device-informed noise and on physical quantum hardware. The main limitations concern activity-threshold sensitivity and the fixed qubit and feature-compression settings. Future work will investigate learned threshold selection, lower-cost circuit designs, improved feature compression, and evaluation across additional quantum devices.

\label{conclusion}

This work introduced A2QTGN, a hybrid quantum--classical model for temporal link prediction. Its main contribution is an adaptive amplitude-encoding mechanism that updates the quantum representation only when meaningful temporal changes are detected.
The experimental results demonstrate that this design improves link prediction across multiple TGBL datasets and provides more reliable generalization than fixed update strategies. The model was also successfully evaluated under device-informed noise and on an IBM quantum processor.
Future work will study learned threshold selection, lower-cost circuit designs, and more scalable feature representations for larger temporal graphs.

\section*{Acknowledgment}
 This work was supported in part by the NYUAD Center for Quantum and Topological Systems (CQTS), funded by Tamkeen under the NYUAD Research Institute grant CG008.

\bibliographystyle{IEEEtran}

{\setstretch{0.95}

{\scriptsize
\bibliography{refs}}}
\end{document}